\documentclass{astro}

\usepackage{epsf}

\newcommand{\md}{\mbox{d}}

\newcommand{\dn}{\rm{dn}}
\newcommand{\sn}{\rm{sn}}

\begin{document}

   \thesaurus{03         
              (02.01.1;  
               02.18.5;  
               13.07.3;  
               11.01.2)} 

   \titlerunning{Particle acceleration in AGN}

   \title{Particle acceleration by rotating magnetospheres in
          active galactic nuclei}

   \author{F.M. Rieger \& K. Mannheim}

   \offprints{F.M. Rieger, frieger@uni-sw.gwdg.de}

   \institute{Universit\"ats-Sternwarte, Geismarlandstr. 11, 
              D-37083 G\"ottingen, Germany}

   \date{Received \dots}

   \maketitle

   \begin{abstract}
      We consider the centrifugal acceleration of charged test particles 
      by rotating magnetospheres which are widely believed to be responsible 
      for the relativistic jet phenomenon in active galactic nuclei (AGN).
      Based on an analysis of forces the equation for the radial accelerated 
      motion is derived and an analytical solution presented under the 
      assumption of an idealized spherical magnetosphere. 
      We show that the rotational energy gain of charged particles moving 
      outwards along rotating magnetic field lines is limited in general by
      (i) inverse-Compton losses in the radiation field of the 
      disk in which the magnetosphere is anchored and 
      (ii) the breakdown of the bead-on-the-wire approximation which 
      occurs in the vicinity of the light cylinder. 
      The corresponding maximum Lorentz factor for electrons is of the order
      of a few hundred for the sub-Eddington conditions regarded to be 
      typical for BL~Lacs. In AGN with enhanced accretion rate the 
      acceleration mechanism seems to be almost inefficient due to 
      increasing inverse-Compton losses.

   \end{abstract}

\section{Introduction}
    The origin of the nonthermal, highly variable emission in active galactic 
    nuclei (AGN) has been widely discussed. Several acceleration mechanisms 
    have been proposed which may explain the observed high energy emission 
    extending up to TeV energies at least in three blazars (Mkn 421, Mkn 501, 
    1ES 2344+514: e.g. Catanese~1999). Fermi-type particle acceleration 
    mechanisms in relativistic jets seem to be very effective but require a 
    seed population of electrons with Lorentz factors of at least $100$. Up to 
    now, it remains a problem to be solved, how this pre-acceleration is 
    achieved (e.g. Kirk, Melrose, Priest~1994).

    Since the pioneering work of Gold in the late 1960s (Gold~1968,~1969),
    centrifugal driven outflow of matter has often been discussed in
    the context of pulsar emission theory (for recent contribution see 
    e.g. Machabeli \& Rogava~1994; Chedia et al.~1996; Gangadhara~1996; 
    Contopoulos et al.~1999). 
    In the case of accreting black hole systems (e.g. AGN) Blandford and 
    Payne (1982) first pointed out that centrifugal driven outflows (jets) 
    from accretion disks are possible, if the poloidal field direction is 
    inclined at an angle less than $60^{\circ}$ to the radial direction. 
    In such models, a rotating magnetosphere could emerge from an accretion 
    disk or the rotating black hole itself (Blandford \& Znajek 1977) 
    initiating a plasma outflow with initially spherical shape until the 
    flow is collimated on a scale of less than a few hundred Schwarzschild 
    radii (e.g. Camenzind~1995,~1996; Fendt~1997). For a rapidly rotating 
    black hole system, the critical angle mentioned above could be as large 
    as $90^{\circ}$ (Cao~1997). 
    
    In magnetohydrodynamical scenarios for the origin of relativistic jets,
    centrifugal acceleration is rather limited, leading to maximum bulk 
    Lorentz factors of the order of $10$ (Camenzind~1989).    
    Nevertheless, it seems quite interesting whether supra-thermal test 
    particles (e.g. from magnetic flares on the accretion disk) could be 
    accelerated to even higher energies by such rotating magnetospheres.   
    Recently, Gangadhara \& Lesch (1997) proposed a model for spinning 
    active galactic nuclei in which charged test particles are accelerated 
    to very high energies by the centrifugal force while moving along 
    rotating magnetic field lines. According to their calculations, the 
    nonthermal X-ray and $\gamma$-ray emission in AGN could arise via the 
    inverse-Compton scattering of UV-photons by centrifugal accelerated 
    electrons. 
 
    In this paper, we reinvestigate the acceleration of charged test particles 
    in an idealized two-dimensional model where the magnetic field rotates 
    rigidly with a fraction of the rotational velocity of the black hole 
    (cf. Fendt~1997). Centrifugal acceleration occurs as a consequence of the 
    bead-on-the-wire motion. A charged particle gains rotational energy as 
    long as it is directed outwards but we show that its energy gain 
    is substantially limited not only by inverse-Compton losses but also by 
    the effects of the relativistic Coriolis force. 
    
    Based on an analysis of forces, the special relativistic equation of 
    motion is derived in Sect.~2. This equation is solved in closed form 
    in Sect.~3. 
    Sect.~4 gives an estimate for the maximum Lorentz factor attainable in 
    the case of AGN. The results are discussed in the context of the 
    particle acceleration problem for rotating AGN jets in Sect.~5.
    
\section{Analysis of forces in a rotating reference frame}
    Usually the motion of a particle along rotating magnetic field
    lines is treated in the bead-on-the-wire approximation where a bead is 
    assumed to follow the rotating field line and experiences centrifugal 
    acceleration (or deceleration) while moving in the outward direction 
    (e.g. Machabeli \& Rogava~1994; Chedia et al.~1996; Cao~1997). 
    This simple approach yields interesting results, though, as we will 
    show further below, such an approximation breaks down in the region 
    near the light cylinder.

    Let us consider the forces acting on a particle in a rotating frame
    of reference (Gangadhara~1996; Gangadhara \& Lesch~1997).
    A particle with rest mass $m$ and charge $q$, which is injected  
    at time $t_0$ and position $r_0$ with initial velocity $v_0$ parallel to 
    the magnetic field line $B_{\rm r}(t_0)$ experiences a centrifugal 
    force in the radial direction given by
    \begin{equation}
    \vec F_{\rm cf}=m\,\gamma\,(\vec \Omega \times \vec r)\times 
    \vec \Omega\,,
    \end{equation} where $\gamma$ is the Lorentz factor of the particle and
    $\vec \Omega= \Omega \, \vec e_z$ is the angular velocity of the field.
    Additionally, there is also a relativistic Coriolis force in the 
    noninertial frame governed by the equation 
    \begin{equation}
    \vec F_{\rm cor}=m\,\left(2\,\gamma \frac{dr}{dt} + r\,
    \frac{d\gamma}{dt}\right)\,(\vec e_{\rm r} \times \vec \Omega)\,,
    \end{equation} which acts as a deviation-force in the azimuthal direction.
    In the inertial rest frame the particle sees the field line bending off 
    from its initial injection position. Hence, it experiences a Lorentz 
    force, which may be written as
    \begin{equation}
    \vec F_{\rm L}=\,q\,(\vec v_{\rm rel} \times \vec B)\,,
    \end{equation}
   where $v_{\rm rel}$ is the relative velocity between the particle and 
   the magnetic field line and where the convention $c=1$ is used. 
   Due to the Lorentz force a charged particle tries to gyrate around the 
   magnetic field line. 
   Initially, the direction of the Lorentz force is perpendicular to the 
   direction of the Coriolis force, but as a particle gyrates, it changes 
   direction and eventually becomes antiparallel to the Coriolis force. 
   Hence one expects that the bead-on-the-wire approximation holds, if 
   the Lorentz force is not balanced by the Coriolis force. 
   In this case, the accelerated motion of the particle's guiding center 
   due to the centrifugal force is given by 
   \begin{equation}\label{radial}
     \gamma \frac{\md^2 r}{\md t^2} +\frac{\md r}{\md t} \,
     \frac{\md \gamma}{\md t} = \gamma\, \Omega^2 \,r\,,
   \end{equation}where $r$ denotes the radial coordinate and 
   $\gamma=1/\sqrt{1-\Omega^2 r^2-\dot r^2}$.
   The bead-on-the-wire motion for the guiding center breaks down, if the 
   Coriolis force exceeds the Lorentz force, i.e. if the following 
   inequality, given by the azimuthal components of the forces, holds:
   \begin{equation}\label{conditio}
      \frac{\md \gamma}{\md t} > \frac{1}{r}
      \left(\frac{B\,q\,v_{\rm{rel}}}{\,m\,\Omega} -2\,\gamma 
      \frac{\md r}{\md t}\right)\,. 
   \end{equation}

\section{Analytic solution for the radial acceleration}
    The general solution of Eq.~(\ref{radial}) can be found using the 
    simple argument that the energy $E$ of the particle in the rotating 
    reference frame is constant. If $E_0$ denotes the energy of the particle 
    in the inertial rest frame, then the energy $E$ in the uniformly rotating 
    frame (angular velocity $\Omega$) is given in the non-relativistic case 
    by: $E = E_0 - m\,\Omega^2\,r^2$ (e.g. Landau \& Lifshitz~1960). The 
    generalisation of this equation to the relativistic case is 
    straightforward and leads to the transformation
    \begin{equation}\label{energy} 
       E =\gamma\,m\,(1 - \Omega^2\,r^2)\,,      
    \end{equation}where $\gamma$ is the Lorentz factor defined above.

    Assume now, that in the general case a particle is injected at time 
    $t=t_0$ and position $r=r_0$ with initial velocity $v=v_0$. 
    Then, from Eq.~(\ref{energy}) it follows that the time-derivative of the
    radial coordinate $r$ may be written as
    \begin{equation}\label{dr}
    \frac{\md  r(t)}{\md t}=\sqrt{(1-\Omega^2 r^2)
    [1-\tilde{m}\,(1-\Omega^2 r^2)]}\,\,, 
    \end{equation} where $\tilde{m}=(1-\Omega^2 r_0^2-v_0^2)/
     (1-\Omega^2 r_0^2)^2$.  
    In the case $r_0=0$, this expression reduces to the equation 
    given in Henriksen \& Rayburn (1971).\\
    The equation for the radial velocity $v_{\rm r}=\md r/\md t$, 
    Eq.~(\ref{dr}), can be solved analytically (Machabeli \& Rogava~1994), 
    yielding 
    \begin{equation}\label{allgem}
    r(t)=\frac{1}{\Omega}\, \rm{cn}(\lambda_0 - \Omega\,t)\,,
    \end{equation} where $\rm{cn}$ is the Jacobian elliptic cosine  
    (Abramowitz \& Stegun~1965, p.569ff), $\lambda_0$ a Legendre 
    elliptic integral of the first kind:
    \begin{equation}
    \lambda_0= \int_0^{\phi_0}\,\frac{\md \theta}{(1-\tilde{m}\,
    \sin\theta)^{1/2}}\,,
    \end{equation} and where $\phi_0$ is defined by $\phi_0={\rm arccos}
    (\Omega\, r_0)$. By using Eq.~(\ref{allgem}), the time-derivative of $r$ 
    can be expressed as
    $\dot r=\dn(\lambda_0 - \Omega\,t)\,\sn(\lambda_0 - \Omega\,t)$.
    Note that the Jacobian elliptic functions $\rm{sn}$ and $\rm{dn}$ 
    are usually defined by the relations $\rm{sn}^2+\rm{cn}^2=1$ and 
    $\tilde{m}\,\rm{sn}^2+\rm{dn}^2=1$.

    Using Eq.~(\ref{dr}), the Lorentz factor may be written as a function
    of the radial coordinate $r$:
    \begin{equation}\label{gamma_r}
    \gamma=\frac{1}{\sqrt{\tilde m}\,(1-\Omega^2 r^2)}\,.
    \end{equation}
    Thus, in terms of Jacobian elliptic functions one gets
    \begin{equation}
    \gamma=\frac{1}{\sqrt{\tilde m}\,[\sn(\lambda_0 - \Omega\,t)]^{2}}\,.
    \end{equation}

\begin{figure}[htb]
    \vspace{0cm}
    \begin{center}
    \epsfxsize8cm 
    \mbox{\epsffile{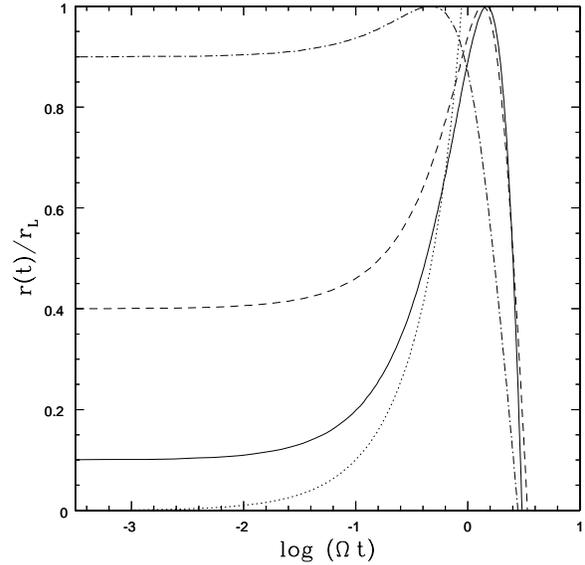}}
    \end{center}
    \vspace{-0.5cm} 
    \caption{\small{The time-dependence of the radial coordinate $r$, plotted 
        for the initial conditions  $v_0=0.99\,c$ and $r_0=0.1\,r_{\rm L}$ 
        (solid line),
        $v_0=0.6\,c$ and $r_0=0.4\,r_L$ (short dashed),  $v_0=0.4\,c$ and 
        $r_0=0.9\,r_L$ (dotted-short dashed); also indicated is the 
        non-relativistic limit: $r(t)\,\Omega/v_0=\rm{sinh}(\Omega\,t)$ 
        (dotted).}}\label{radius}
   \end{figure}

    For the particular conditions where the injection of a test particle is 
    described by $r(t_0=0)=0$ and $v(t_0=0)=v_0$, the time-dependence of the 
    radial coordinate is given by a much simpler expression: In this 
    situation, $\lambda_0$ reduces to a complete elliptic integral of the 
    first kind and therefore, after a change of the arguments, the 
    time-dependence of the radial coordinate becomes
    \begin{equation} 
    r(t)=\frac{v_0\,\rm{sn}(\Omega\,t)}{\Omega\,\rm{dn}(\Omega\,t)}\,.
    \end{equation}
    If one considers non-relativistic motions, where $\tilde{m}\simeq 1$, and
    the special case $r_0=0$, Eq.~(\ref{allgem}) reduces to (cf. 
    Abramowitz \& Stegun~1965) 
    \begin{equation}
     r(t)=v_0\,{\rm{tanh}}(\Omega\,t)/\Omega\, {\rm{sech}}
           (\Omega\,t)=v_0\,{\rm{sinh}}(\Omega\,t)/\Omega\,.
    \end{equation} 
    This expression is known to be the general solution of the equation 
    \begin{equation}
    \ddot r-\Omega^2\,r=0\,,
    \end{equation} which describes the motion of a particle due to the 
    centrifugal force in the non-relativistic limit. In Fig.~\ref{radius}, 
    we compute the time-dependence of the radial coordinate $r$ for different 
    initial conditions under the (unphysical) assumption that the 
    bead-on-the-wire motion continues until the light cylinder (with radius 
    $r_{\rm L}$) is reached. 
    Note that in the relativistic case all particles would turn back at the 
    light cylinder due to the reversal of the centrifugal acceleration 
    (e.g. Machabeli \& Rogava~1994).

\begin{figure}[htb]
    \vspace{0cm}
    \begin{center}
    \epsfxsize8cm 
     \mbox{\epsffile{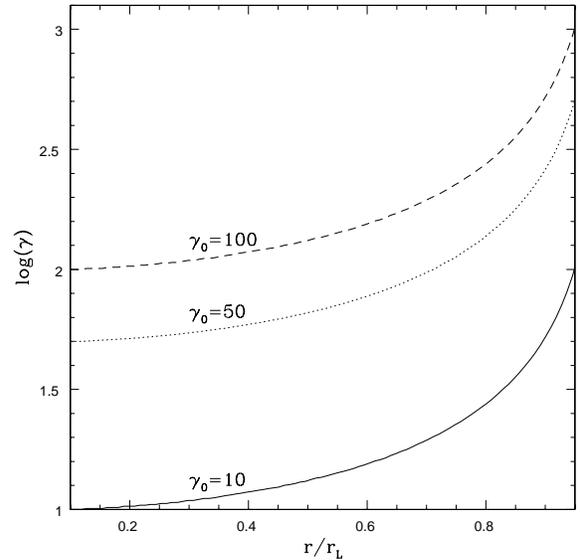}}
       \end{center}
       \vspace{0.0cm} 
       \caption{\small{The relativistic Lorentz factor $\gamma$ for a particle 
        approaching the light cylinder $r_{\rm L}$ using $r_0=r_{\rm L}/10$ 
        and injection Lorentz factors $\gamma_0=10\,$ (solid line), 
        $\gamma_0=50\,$ (dotted) and $\gamma_0=100\,$ (dashed).}}\label{gamma}
   \end{figure}

\begin{figure}[htb]
       \vspace{-0.5cm}
       \begin{center}
         \epsfxsize8cm 
         \mbox{\epsffile{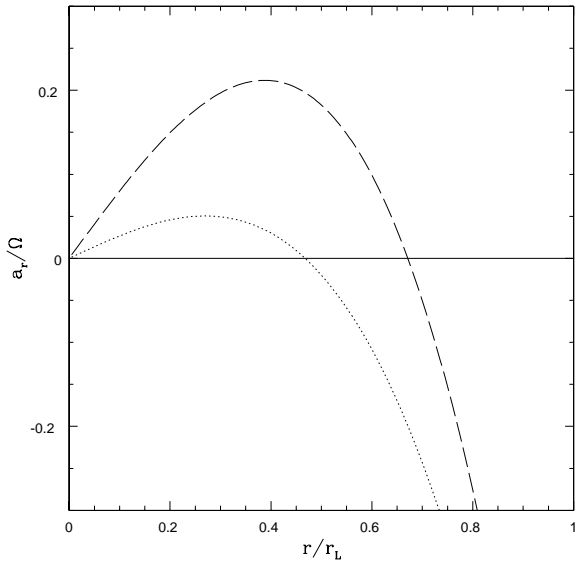}}
       \end{center}
       \vspace{-0.5cm} 
       \caption{\small{The radial acceleration $a_{\rm r}$ as a function of 
        $r/r_{\rm L}$ for the initial conditions $r_0=0$ and $v_0=0.3\,c$ 
        (dashed), $v_0=0.6\,$c (dotted).}}\label{accel}
   \end{figure}

    Using the definition of the Lorentz factor, the equation for the 
    accelerated motion, Eq.~(\ref{radial}), may also be written as
    \begin{equation}\label{centri}
    \frac{\md^2 r}{\md t^2}=\frac{\Omega^2 r}{1-\Omega^2\,r^2}
    \left[1-\Omega^2\,r^2 - 2\, \left(\frac{\md r}{\md t}\right)^2\,
    \right]\,.
    \end{equation}
    (cf. Chedia et al.~1996; Kahniashvili et al.~1997). By inserting the 
    above relations, the solution for the radial acceleration could be 
    expressed in terms of the Jacobian elliptic functions: 
    \begin{equation}
    \ddot{r}=\Omega\,\cdot \rm{cn}(\lambda_0 - \Omega\,t)\,
    [1 - 2\,\rm{dn}^2(\lambda_0 - \Omega\,t)].
    \end{equation}
    According to our simple model, one expects that a charged test particle 
    gains energy due to rotational motion as long as it is directed outwards. 
    Therefore the relativistic Lorentz factor increases with distance $r$ 
    as the particle approaches the light cylinder. 
    This is illustrated in Fig.~\ref{gamma} where we plot the evolution of the 
    relativistic Lorentz factor $\gamma$ as a function of $r$ for different 
    initial velocities $v_0$ and fixed $r_0=r_{\rm L}/10$ (using a typical 
    light cylinder radius of $r_{\rm L} \simeq 10^{15}$cm).
    Note that $\gamma(r/r_{\rm L})$ is not scale-invariant with respect to the 
    injection velocity $v_0$ (i.e. the injection energy).
    If one identifies Eq.~(\ref{centri}) with the general expression for the
    centrifugal force, which reduces in the non-relativistic limit to the 
    well-known classical expression, the centrifugal force changes its signs 
    and becomes negative for $r^2/r_{\rm L}^2>1-(2\,\tilde{m})^{-1}$ 
    (see Fig.~\ref{accel}; cf. also Machabeli \& Rogava~1994). Hence, 
    if one assumes that the bead-on-the-wire approximation holds in the 
    vicinity of the light cylinder, the radial velocity becomes zero at the 
    light cylinder and changes direction in any case. Accordingly, a crossing
    of the light cylinder within the bead-on-the-wire approximation, as
    mentioned for example in Gangadhara \& Lesch~1997, is not physical 
    (cf. Fig.~\ref{radius}).
    The reversal of the direction of centrifugal force according to which 
    the centrifugal force may attract rotating matter towards the centre is 
    well-known in strong gravitational fields (for Schwarzschild geometry: 
    Abramowicz~1990; Abramowicz \& Prasanna~1990; for Kerr geometry:
    Iyer \& Prasanna~1993; Sonego \& Massar~1996).
    For illustration, we compute in Fig.~\ref{accel} the evolution of the 
    effective radial acceleration $a_{\rm r}=\md^2r/\md t^2$ as a function 
    of the radial coordinate $r$ for different initial velocities. Obviously, 
    there exists a point where the effective acceleration, i.e. the 
    centrifugal force, becomes negative. 

\section{Estimate of the maximum Lorentz factor}    
    Consider now the acceleration of electrons via rotating magnetospheres 
    in AGN. Imagine an electron which moves along a rotating magnetic field 
    line towards the light cylinder. Generally, one expects that there are 
    two processes which could limit the energy gain of a particle:

    First, there are inverse-Compton energy losses due to interaction with
    accretion disk photons: low energy accretion disk photons are scattered 
    to higher energies by the accelerated electrons so that the photons gain 
    energy while the electrons lose energy. Near the disk the electrons might
    encounter a very strong disk radiation field, which substantially limits  
    the maximum attainable energy (this need not be the case if electrons 
    are accelerated far away from the disk, e.g. Bednarek, Kirk \& 
    Mastichiadis~1996). The maximum energy, which an electron is able to 
    reach under the influence of inverse-Compton scattering is given at 
    the point where the acceleration time scale equals the cooling time scale.
    In the case, where the energy of the photon in the electron rest frame is 
    small compared to the energy of the electron (Thomson scattering), the 
    cooling time scale for inverse-Compton losses can be approximated by 
    (e.g. Rybicki \& Lightman~1979)
    \begin{equation}\label{inverse}
      t_{\rm cool}^{\rm IC} = 3\times 10^7\,\frac{\gamma}{(\gamma^2-1)\, 
      U_{\rm rad}}\,\mbox{\rm [s]}\,,
    \end{equation} where $U_{\rm rad}=\tau\,L_{\rm disk}/4\,\pi\,r_{\rm L}^2$ 
    is the energy density of the disk radiation field and $\tau \leq 1$.\\
    If one uses Eq.~(\ref{gamma_r}), the acceleration time scale 
    $t_{\rm acc}$ may be written as:
    \begin{equation}\label{acc}
      t_{\rm acc}= \gamma/\dot \gamma = \frac{\sqrt{1-\Omega^2\,r^2}}
      {2\,\Omega^2\,r\,\sqrt{1-\tilde m\,(1-\Omega^2\,r^2)}}\,.
    \end{equation} 
    By equating this two time scales we obtain an estimate for the 
    maximum electron Lorentz factor $\gamma_{\rm max}$.

    A second, general constraint, which was neither considered by Machabeli \& 
    Rogava~(1997) nor used in the calculation by Gangadhara \& Lesch~(1997), 
    is given by the breakdown of the bead-on-the-wire approximation which 
    occurs in the vicinity of the light cylinder. Beyond this point, where 
    the Coriolis force exceeds the Lorentz force [see condition 
    Eq.~(\ref{conditio})], the particle leaves the magnetic field line 
    so that the rotational energy gain ceases. Hence the acceleration 
    mechanism becomes ineffective. In the case of AGN, where the magnetic 
    field strength is much smaller than in pulsars, this constraint may be 
    quite important.

    For illustration, we apply our calculations in the following to a typical 
    AGN using a central black hole mass $M_{\rm BH} = m_8\,10^8\,M_{\sun}$ 
    and a light cylinder radius $r_{\rm L} \simeq 10^{15} m_8$~cm, 
    where $M_{\odot}$ denotes the solar mass. 
    The Eddington luminosity, i.e. the maximum luminosity of a source of 
    mass $M_{\rm BH}$ which is powered by spherical accretion, is given by
    $L_{\rm Edd} \simeq 10^{46}\,{\rm ergs\,\,s}^{-1}$. Typically, we may 
    express the disk luminosity as $L_{\rm disk}=l_{\rm e} \times 
    L_{\rm Edd}$, with $10^{-4}< l_{\rm e} \leq 1$. 
    Thus, the equipartition magnetic field strength at the radius $r$ is 
    given by $B(r)^2=2\,L_{\rm disk}/r^2$.
    Electrons are assumed to be injected at an initial position $r_0\simeq 
    0.4\,r_{\rm L}$ with a characteristic escape velocity from the last 
    marginally stable orbit around a black hole of $v_0\simeq 0.6\,$c.
    By applying the two constraints above, we get three generic regimes 
    for the acceleration of electrons by rotating magnetospheres:
    \begin{enumerate}
     
    \item the region, in which inverse-Compton losses dominate entirely over 
    the energy gains, leading to an inefficient acceleration (generally in the 
    case of Eddington accretion, i.e. $l_{\rm e} \sim 1$). 
 
    \item the region, in which inverse-Compton losses are important but not 
    dominant (generally the sub-Eddington range: $l_{\rm e} \leq 
    2\times10^{-2}$).
    In this case the acceleration mechanism works, but there exists a maximum 
    Lorentz factor given at the position where the energy gain is exactly 
    balanced by losses. This is illustrated in Fig.~\ref{cooling}, where
    we calculate the cooling and the acceleration time scale as a function of
    the Lorentz factor $\gamma$ for $l_{\rm e}=5\times 10^{-3}$.
    For this value, the maximum Lorentz factor is roughly 
    $\gamma \simeq 150$. Typically, the maximum Lorentz factors in this 
    range are of the order of $100$ to $1000$ (see Fig.~\ref{second}).
   \begin{figure}[htb]
   \vspace{-0.5cm}
   \begin{center}
   \epsfxsize8cm 
  \mbox{\epsffile{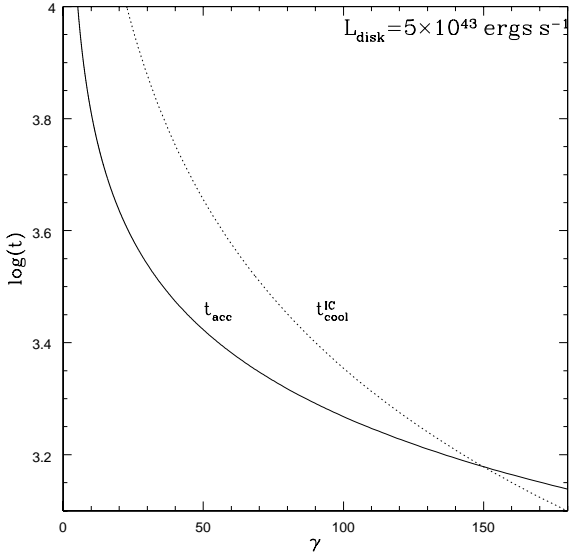}}
   \end{center}
   \vspace{-0.5cm} 
   \caption{\small{Cooling times scale $t_{\rm cool}^{\rm IC}$ for 
       inverse-Compton scattering, Eq.~(\ref{inverse}), and acceleration 
       time scale $t_{\rm acc}$, Eq.~(\ref{acc}), as a function of the Lorentz
       factor $\gamma$ using $l_{\rm e}=5\times 10^{-3}$ and $\tau=1$. The 
       maximum electron Lorentz factor, given at the position where the 
       cooling time scale equals the accelerations time scale, is 
       approximately $150$.}} \label{cooling}
   \end{figure}
   \begin{figure}[htb]
   \vspace{0cm}
   \begin{center}
   \epsfxsize8cm 
  \mbox{\epsffile{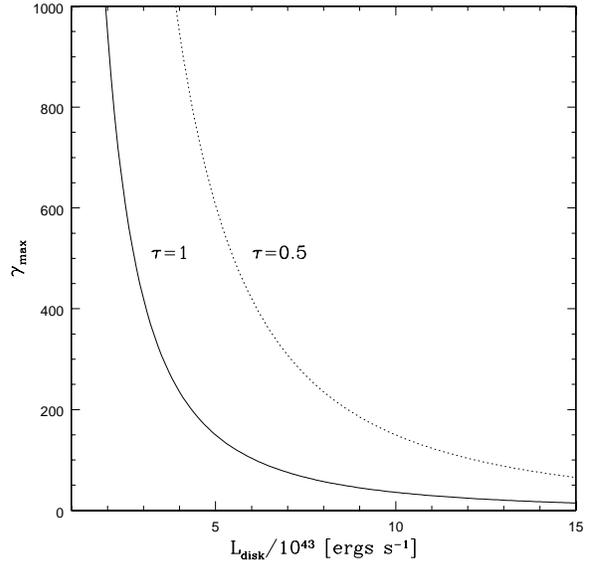}}
   \end{center}
   \vspace{-0.5cm} 
  \caption{\small{Maximum electron Lorentz factor $\gamma_{\rm max}$ attainable
         under the influence of inverse-Compton losses as a function of the 
         disk luminosity $L_{\rm disk}$ for $\tau=0.5$ (dotted) and 
         $\tau=1$ (solid), where $\tau=4\,\pi\,r_{\rm L}^2 U_{\rm rad}/
         L_{\rm disk}$ and $U_{\rm rad}$ being the energy density of the 
         disk radiation field.}}\label{second}
   \end{figure}
   \item the region, in which the inverse-Compton losses are rather 
   unimportant (generally $l_{\rm e} < 10^{-3}$).
   In this case, the maximum Lorentz factor is determined by the breakdown of 
   the bead-on-the-wire approximation [see Eq.~(\ref{conditio})], which yields 
   a general upper limit for the Lorentz factor of the order of $1000$. This 
   limit is found if one approximates $v_{\rm rel}$ by the light velocity 
   which amounts to the highest value for the Lorentz forces. The results 
   are shown in Fig.~\ref{third}, where we also allow the injection position 
   to vary. 
   We wish to note, that the results, presented in Fig.~\ref{third}, depend 
   essentially on the assumed intrinsic magnetic field strength and the size 
   of the light cylinder radius (i.e. the angular velocity). 
   Generally, for a sufficient approximation, the maximum Lorentz factor  
   is given by:
   \begin{equation} 
   \gamma_{\rm max} \simeq 
   \frac{1}{\tilde{m}^{1/6}}\,\left(\frac{B(r_{\rm L}) \,q}
   {2\,m\,\Omega\,c}\right)^{2/3}\,.
   \end{equation}
   Thus, even if one uses a magnetic field strength of $B(r_{\rm L})= 100\,$G, 
   which is roughly three times the corresponding equipartition field, 
   the maximum Lorentz factor does not exceed $2.5\times 10^3$. 
   \begin{figure}[htb]
   \vspace{0cm}
   \begin{center}
   \epsfxsize8cm       
   \mbox{\epsffile{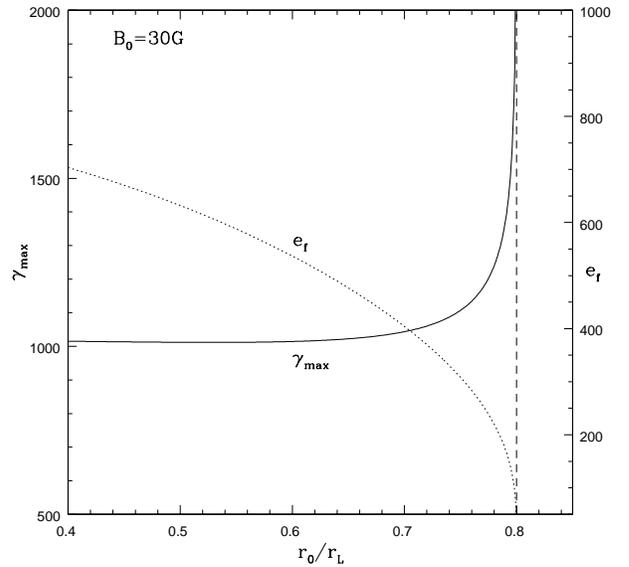}}
   \end{center}
   \vspace{-0.5cm} 
   \caption{\small{Maximum electron Lorentz factor $\gamma_{\rm max}$ as a 
      function of the initial injection position $r_0$ for $v_0=0.6$ c and 
      $B(r_{\rm L})=30\,$G (i.e. a disk luminosity $L_{\rm disk} \simeq1.35
      \times 10^{43}{\rm ergs\,\,s}^{-1}$). The dotted line shows the 
      decrease in efficiency of energy gain $e_f=\gamma_{\rm max}/\gamma_0$, 
      while the dashed line indicate the relativistic limit for injection 
      given by the condition $1-v_0^2-\Omega^2\,r_0^2 > 0$.}}\label{third}
   \end{figure}
   \end{enumerate}

\section{Discussion}
    We have considered the acceleration of charged test particles via 
    rotating magnetospheres based on a model topology which is motivated 
    by the standard model for AGN (cf. Begelman~1994; Camenzind~1995; 
    Fendt~1997). Accordingly, the jet magnetosphere originates very close 
    to the central black hole from an accretion disk, with initially spherical
    profile until the relativistic jet is collimated to a cylindrical shape 
    outside the light cylinder.

    The centrifugal particle acceleration model described in this 
    paper extends the calculations by Machabeli \& Rogava~(1994) and 
    Gangadhara \& Lesch~(1997). We find that the maximum Lorentz 
    factor attainable for an electron moving along a rotating magnetic
    field line is substantially limited not only by radiation losses 
    (e.g. inverse-Compton) but also by the breakdown of the bead-on-the-wire 
    approximation which occurs in the vicinity of the light cylinder. 
    Due to these limiting effects, the acceleration of particles by rotating 
    magnetospheres seems to be rather less important in the case of AGN. 
    Our current calculations indicate, that for sub-Eddington accreting black 
    holes, such as black holes with advection-dominated accretion flows 
    (e.g. Narayan \& Yi~1994; Narayan~1997), efficient pre-acceleration of 
    electrons to Lorentz factors of the order of a few hundred might be 
    possible, at least under the highly idealized conditions of our 
    analytical toy model. It seems interesting that the highest energy 
    gamma rays have been discovered from AGN of the BL Lac type which very 
    likely accrete in a sub-Eddington mode (e.g. Celotti, Fabian \& Rees~1998).
    Under such conditions, inverse-Compton scattering of accretions disk 
    photons with energy $\sim 0.1$ keV produces gamma rays with a maximum 
    energy of $\sim 100~(\gamma_{\rm max}/10^3)^2$~MeV, which is in general 
    too low to explain the observed high-energy gamma rays in blazars 
    (e.g. Kanbach~1996; Catanese~1999).
 
    We wish to mention that the results in this paper essentially depend on 
    the assumed intrinsic magnetic field and the angular frequency $\Omega$, 
    i.e. on the size of the light cylinder radius ($\gamma_{\rm max} \propto 
    (B/\Omega)^{2/3}$). 
    Therefore, one could find a way out of the problem above, for 
    example, by assuming a light cylinder radius in BL Lac type objects 
    which is much greater than $10^{15}\,\mbox{\rm cm}$ for a black hole 
    mass of $M_{\rm BH} = 10^8\,M_{\sun}$. However, in view of 
    magnetohydrodynamic models already existing, this seems to be rather 
    improbable. 
    In any case, the acceleration of supra-thermal test particles by rotating 
    magnetospheres might possibly provide an interesting explanation for the 
    pre-acceleration which is required for efficient Fermi-type particle 
    acceleration at larger scales in radio jets.
   
    There are several restrictions on our approach, e.g. we have assumed 
    a projected, two-dimensional geometry and rigid rotation of magnetic field 
    lines almost up to the light cylinder, hence, concerning the last point,
    neglected a kind of toroidal twist (Begelman~1994), when the inertial 
    forces overcome the tension in the field line so that the field line is 
    swept back opposite to the sense of rotation. However, one would not expect
    that these restrictions alter our conclusions essentially since they 
    should lower the upper limit for the maximum Lorentz factor by making the 
    acceleration mechanism ineffective somewhat earlier. Another restriction 
    is the use of special relativity in our analysis, which is only justified 
    far away from the black hole. A detailed general relativistic model is 
    needed to assess whether this might affect the results very strongly or 
    not.

\begin{acknowledgements} We would like to thank Harald Lesch for comments.
      F.M.R. acknowledge support under DFG Ma 1545/2-1.
\end{acknowledgements}

\end{document}